\documentclass{emulateapj}

\usepackage{times}

\newcommand{\teff}{$T_{\rm eff}$}
\newcommand{\feh}{\mbox{[M/H]}}

\newcommand{\vsini}{\mbox{$v\,\sin\,i$}}
\newcommand{\vrad}{\mbox{$v_{\rm rad}$}}
\newcommand{\vmicro}{\mbox{$\xi$}}
\newcommand{\vmacro}{\mbox{$\Xi$}}
\newcommand{\kms}{\mbox{km s$^{-1}$}}

\submitted{To appear in the Astrophysical Journal}
\shorttitle{Metallicity of the HD 98800 System}

\begin{document}
\title{The Metallicity of the HD 98800 System}
\author{Tanmoy Laskar$^{1,2}$, David R. Soderblom$^2$, Jeff A. Valenti$^2$, and John R. Stauffer$^3$}
\affil{Cavendish Laboratory, University of Cambridge, JJ Thomson Avenue, Cambridge CB3~0HE, UK; tl291@cam.ac.uk\\
Space Telescope Science Institute, 3700 San Martin Dr., Baltimore MD 21218; soderblom@stsci.edu, valenti@stsci.edu\\
Spitzer Science Center, California Institute of Technology, Pasadena CA 91125; stauffer@ipac.caltech.edu}





\begin{abstract}
Pre-main-sequence (PMS) binaries and multiples enable critical tests of stellar models if masses, metallicities, and luminosities of the component stars are known. We have analyzed high-resolution, high signal-to-noise echelle spectra of the quadruple-star system HD 98800 and using spectrum synthesis, computed fits to the composite spectrum for a full range of plausible stellar parameters for the components. We consistently find that sub-solar metallicity yields fits with lower $\chi^2$ values, with an overall best-fit of $\feh = -0.20\pm0.10$. This metallicity appears to be consistent with PMS evolutionary tracks for the measured masses and luminosities of the components of HD 98800 but additional constraints on the system and modelling are needed. 
\end{abstract}

\objectname[HD 98800]{}

\section{The HD 98800 System}
HD 98800 is a multiple system of pre-main-sequence (PMS) stars that is particularly well-suited to testing evolutionary models.  It has two visual components \citep{Innes1909}, separated by $0.848$ arcsec in 2001 \citep{Prato2001} and estimated to have an orbital period of about 300 to 430 years \citep{Tokovinin1999}. In addition, the brighter (southern) `A' component is a single-lined spectroscopic binary with a period of 262 days, while the fainter `B' component is a double-lined spectroscopic binary with a period of 315 days \citep{Torres1995,Boden2005}. Thus there are at least four stars in the system, designated as Aa, Ab, Ba, and Bb. Of these, the light from only three of the stars, Aa, Ba, and Bb, is visible in a combined optical spectrum \citep{Soderblom1998}.

\citet{Prato2001} fitted the estimated luminosity and effective temperature of Aa with pre-main sequence evolutionary models and determined its mass to be $1.1\pm0.1 M_\sun$. \citet{Boden2005} measured the orbit of the B-subsystem and determined masses of Ba and Bb as $0.699\pm0.064$ and $0.582\pm0.051$ $M_\sun$ respectively, by using Keck Interferometer observations combined with Hubble Space Telescope (\textit{HST}) astrometry and radial velocities from \citet{Torres1995}. \textit{Hipparcos} measured a distance of $46.7\pm 6.2$ pc \citep{ESA1997b}. Using their orbital solution, \citet{Boden2005} derived an independent distance estimate, placing the system at $42.2 \pm 4.7$ pc. A new reduction of the Hipparcos data by \citet{Leeuwen2007a} gives a parallax of $22.28 \pm 2.3$ arcsec \citep[see][]{Leeuwen2007}, corresponding to a distance of $44.9 \pm 4.6$ pc. Based on the lithium abundances of the components, \citet{Soderblom1998} estimated the age of HD 98800 to be $10\pm3$ Myr. In addition, HD 98800 is a member of the TW Hya association \citep{Kastner1997}.

With well-determined masses and parallax, HD 98800 provides a unique opportunity for testing PMS evolutionary models. Not only are the masses and luminosities constrained for HD 98800, but having several stars in a single system helps to distinguish between stellar evolutionary models. This is because we assume that the stars in this system are of the same age and composition and hence should lie near a single isochrone in a Hertzsprung$-$Russell (HR) diagram. The available model isochrones for low mass stars differ from one another significantly, primarily because of the different ways the models treat convection.  Until now it has not been possible to distinguish between the models with any certainty because the metallicity of the system was unknown. Further, based on the locations of the Ba and Bb components in an HR diagram, \citet{Boden2005} showed that the PMS evolutionary models of \citet{Siess2000} and \citet{Baraffe1998} yielded masses inconsistent with their dynamical masses, unless HD 98800 has significantly subsolar metallicity of about $-0.3$ to $-0.5$ dex.

One might be tempted to adopt for the metallicity of HD 98800, a value derived from another member of the TW Hya association. \citet{Yang2005} estimated the metallicity of TW Hya itself, but the result ($\feh=-0.10\pm0.12$) was imprecise since TW Hya is an accreting T Tauri star \citep{Muzerolle2000,Alencar2002}, which added uncertainty to the spectrum analysis. Other members of the association are much cooler and the available molecular line data, required for low temperature atmospheres, are inadequate for precise abundance analyses. Aside from multiplicity, HD 98800 is a useful system for an abundance analysis because its components are warm enough to have spectra similar to the Sun's (i.e., they do not have significant molecular features).

In this paper we put observational constraints on the metallicity of the HD 98800 system from analysis of high resolution and high signal-to-noise spectra. Our analysis is complicated by that which makes HD 98800 so interesting: the multiplicity of the system.  Components A and B are not far enough apart to observe separately from the ground without the aid of adaptive optics, and so the spectrum observed consists of three stellar spectra added together. 
The composite nature of the spectrum means that our ability to determine the abundances is lessened because continuum emission from all the companions dilutes the depths of the spectrum features.  We circumvent this by fitting synthetic spectrum models for a variety of temperatures and radii and find that the best fit models for the system indicate subsolar metallicity. 

\begin{figure}
\plotone{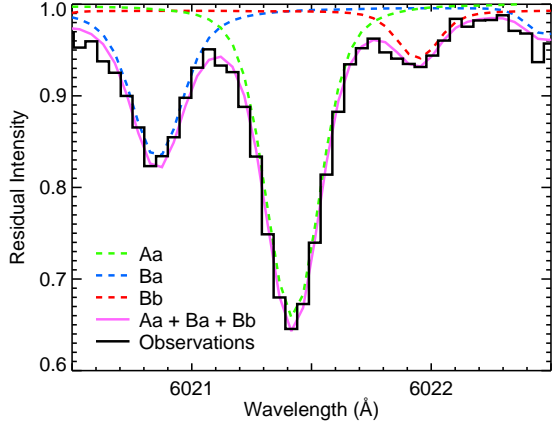}
\caption{Observed spectrum (histogram) of HD 98800 in the vicinity of the isolated \ion{Mn}{1} 6021.793 \AA\ line. The three components in the observed spectrum correspond to contributions (dashed curves) from Ba (blue-shifted), Aa (central), and Bb (red-shifted). The sum of the components (smooth curve) matches the observations well. \label{fig_BBplot}}
\end{figure}

\begin{deluxetable}{c c c c c}
\tablewidth{0pt}
\tablecaption{Observation Log \label{Obstable}}
\tablehead{
\colhead{Night} & \colhead{HJD} & \colhead{Exposure} & \colhead{H.A.} & \colhead{S/N}\\
\colhead{ } & \colhead{ } & \colhead{(s)} & \colhead{ } & \colhead{ }
}
\startdata
I   & 2454075.062 & 1500 & $-01:12:47.7$ & 95-120\\
II  & 2454076.060 & 1500 & $-01:12:43.1$ & 120-160\\
III & 2454077.067 & 1500 & $-00:58:12.2$ & 110-150\\
\enddata
\end{deluxetable}

\section{Observations}
\label{Observations}
We obtained spectra of the HD 98800 star system on three nights in 2006 December (Table \ref{Obstable}) with the Hamilton echelle spectrograph at Lick Observatory \citep{Vogt1987}, when the spectral lines of the Aa, Ba and Bb components were visible and well separated in radial velocity (see Figure \ref{fig_BBplot}). The data were reduced using standard IRAF procedures in the \textit{echelle} package. The resolving power of the instrument was about 40,000. The spectra have a signal-to-noise ratio per pixel of over 100 on each night of observation and cover wavelengths from $3920$ \AA\ to $9920$ \AA. 

At the time of observation the A and B components were separated by 0.848 arcsec at position angle $4.4\,^{\circ}$  and the system was observed at hour angles near 1 hour east, meaning that A and B were separated by 0.3 arcsec along the dispersion direction at the slit. A very wide slit 
(1.2 arcsec)
 was used to facilitate the recording of the full light from both the visual components. Despite this, we found that the relative line depths of Aa to Ba+Bb were not constant from one night to the next, which we interpret as a variation in the placement of A and B in the slit from night to night. This effect adds uncertainty to the Aa/(Ba + Bb) flux ratio, as we note below. However, it should not affect the Ba/Bb flux ratio since the Ba/Bb separation is negligible compared to the slit width and the B subsystem is not resolved.

\section{Spectrum Analysis}
\subsection{Synthetic spectra}
We fitted the observed spectra by computing separate synthetic spectra for each of the three stars using Spectroscopy Made Easy (SME), described in \citet{Valenti1996}. SME is a set of IDL routines coupled to a C++ library that contains modules for computing opacities and elemental ionization fractions as well as radiative transfer. The radiative transfer code assumes local thermodynamic equilibrium throughout the line-forming depths of the stellar atmosphere.

Model atmospheres for the spectrum synthesis \citep{Kurucz1992} in SME are taken from \cite{Kurucz1993a}, who provides opacities on a grid of various stellar parameters. The method of interpolating spectra (or model atmospheres, as SME does) to find the best fit requires knowledge of the star's effective temperature (\teff), surface gravity ($\log g$), metallicity (\feh), microturbulence (\vmicro), radial-tangential macroturbulence (\vmacro), projected rotational velocity (\vsini) and radial velocity (\vrad). Here \feh\ is a scale factor applied to the solar elemental abundances to generate stellar abundances. 

Throughout our analysis, we used the $\log g$ values for the Ba and Bb components from \citet{Boden2005} (henceforth B05), empirical \vsini\ values for Aa and Ba from \citet{Soderblom1996} and nominal values of 4.25 for the gravity of Aa and 0.0 for
 the \vsini\ of Bb (Table \ref{98800params1}). Given our resolution and signal-to-noise, we cannot distinguish \vsini\ for Bb
 from zero. We took \vmicro\ to be 0.85 \kms\ from \citet{Valenti2005} (henceforth VF05) and used VF05's relation between
 \vmacro\ and \teff. While previous studies \citep{Padgett1996,James2006,Santos2008} have found microturbulent velocities
 in T Tauri stars in the 2-4 \kms\ range, HD98800 is only slightly above the main sequence \citep{Soderblom1996} and is
 not nearly as active as most classical T Tauri stars. A higher value for \vmicro\ than 0.85 would force our derived
 metallicity to be even lower than we find it, which as we will show below, would only exacerbate the already low \feh\
 that we find for this system.

Using the above parameters, we synthesized spectra for each of the three stars in the HD 98800 spectrum, then shifted each synthesized spectrum by the observed radial velocities (Section \ref{vradtext}) of the stellar components and co-added the three shifted synthetic spectra in proportion to the flux ratios (Section \ref{fluxratios}) of the stars to obtain the final modeled spectrum.

\begin{deluxetable}{c c c c c c}
\tablewidth{0pt}
\tablecaption{Input parameters for radial velocity estimates\label{98800params1}}
\tablehead{
\colhead{Component} & \colhead{$\log g$} & \colhead{\vsini} & \colhead{\teff}
 & \colhead{$R/R_\sun$}  & \colhead{$c_i$}\\
\colhead{ } & \colhead{ } & \colhead{(\kms)} & \colhead{(K)}
 & \colhead{ }  & \colhead{ }
}
\startdata
Aa & 4.25 & 5.0 & 4500 & 1.75 & 0.69\\
Ba & 4.21 & 3.0 & 4050 & 1.09 & 0.24\\
Bb & 4.34 & 0.0 & 3550 & 0.63 & 0.07\\
\enddata
\end{deluxetable}

\subsection{Computed flux ratios}
\label{fluxratios}

Flux ratios for the three stars, which enter into the above step, were computed by taking the Planck function, $B(\lambda,T)$ at the center of our wavelength range at 6115 \AA{} and using initial estimates of the radii, $R_i$ of the stars from B05. 
We have for the fractional flux contribution from any one star in a spectroscopic multiple system of $N$ stars

\begin{eqnarray}
 c_i(\lambda,T_i) 
= \frac{R_i^2 B(\lambda,T_i)} {\sum_{j=1}^{N}{R_j^2 B(\lambda,T_i)}} \label{fluxratioeqn1}
\end{eqnarray}
\noindent
with $N=3$ in our case. The combined spectrum modeled as seen from the Earth is then given by

\begin{eqnarray}
 S_{T} = \sum{c_j(\lambda,T_j) \times f_j(\lambda_{0},v_{rad_{j}},(\vsini)_j)}
\end{eqnarray}
\noindent
where $f_j$ is the intrinsic spectrum of star $j$ (normalized to unity) at the rest wavelength $\lambda_0$ and is a function of the star's radial velocity $(v_{rad})_{j}$ and projected rotation velocity $(\vsini)_j$. An example of models for the individual stars weighted by their flux ratios and appropriately shifted in radial velocity, and the combined spectrum are shown along with the corresponding segment of the observations in Figure \ref{fig_BBplot}.

\begin{deluxetable}{c r c r c r c}
\tablewidth{0pt}
\tablecaption{Radial velocity estimates (\kms) \label{vradfit}}
\tablehead{
\colhead{Component} & \multicolumn{2}{c}{Night I} & \multicolumn{2}{c}{Night II}
 & \multicolumn{2}{c}{Night III}\\
\cline{2-7}
\colhead{} & \colhead{obs.} & \colhead{$\Delta$\tablenotemark{a}} & \colhead{obs.} & 
\colhead{$\Delta$\tablenotemark{a}} & \colhead{obs.} & \colhead{$\Delta$\tablenotemark{a}}
}
\startdata
Aa &      $8.80$ &  $0.42$ &     $7.38$ & $1.89$ &   $7.42$ & $1.90$ \\
Ba & $-20.60$ & $-1.95$ & $-20.82$ & $-1.35$ & $-20.28$ & $-1.35$ \\
Bb &  $36.42$  & $3.24$ &  $33.58$ & $5.61$ &  $34.72$ & $3.83$ \\
\enddata
\tablenotetext{a}{Difference (model $-$ observed) between model radial velocities \citep{Torres1995} and measured values}
\end{deluxetable} 

\subsection{Determination of radial velocities and an improved ephemeris}
\label{vradtext}
The radial velocities we measure are necessary input to the code as part of the fitting process. We first determined the radial velocities of the three stellar components by taking nominal values of the temperatures and radii for the stars and followed the synthesis procedure outlined above. Table \ref{98800params1} shows the parameters of the three stars used to generate spectra for radial velocity fitting. These values are close to the final values obtained after minimizing the $\chi^2$ statistic between the observed and modeled spectra (Section \ref{abunds}).

These radial velocity estimates were determined from visual inspection, which was adequate to match the models to the observed spectrum to determine metallicity. We did not base our radial velocity estimates on the $\chi^2$ values for the fits since the final mask (Section \ref{matching}) for the combined spectrum changes with the radial velocity parameters for the three stars, introducing a bias in the behavior of $\chi^2$. The $\chi^2$ values between different radial velocities are not comparable and hence not good estimators of when the radial velocities have been best determined. However, based on visual examination, differences in \vrad\ of 0.2 \kms\ for Aa, 0.5 \kms\ for Ba and 1.5 \kms\ for Bb were easily discernible and we take these as conservative estimates of our measurement precision.

\begin{figure}
\caption{Observed minus predicted radial velocities of each component (Aa, Ba, Bb) of HD 98800 and each observing night (I,II,III) assuming the nominal ephemeris \label{vradfitfig1}}
\plotone{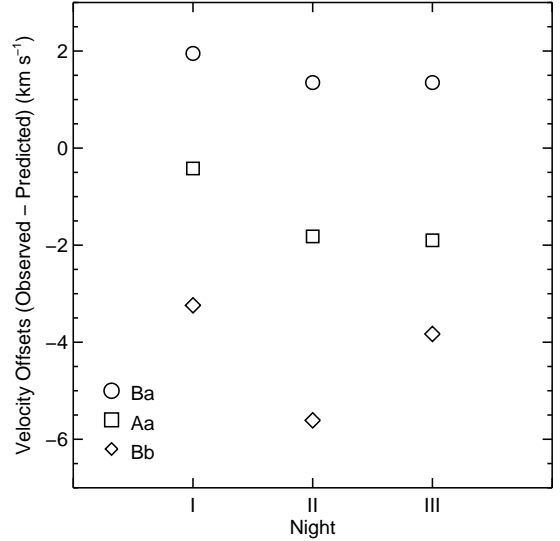}
\end{figure}

Table \ref{vradfit} lists the measured radial velocities, along with a comparison to the modeled orbital curve from \cite{Torres1995} (henceforth T95), in the sense of model minus observed. Our measured radial velocities differ significantly from the predicted values (Figure \ref{vradfitfig1}). As noted above (Section \ref{Observations}), we used a wide slit in order to minimize slit losses. This allowed the placement of the stars in the slit to differ between nights, causing the zero-point velocity to change. A shift of the order of 
$50\%$  
 of the slit width would correspond to movement of the spectrum by 1 pixel, or 2.5 \kms\ at the detector. Further, the spectrum on the CCD shifts vertically, depending on the zenith angle of the object to which the telescope is pointed. This introduces errors into the wavelength calibration, since the fixed Th-Ar lamp utilized for this purpose would form spectral features on the detector at the same pixels, independent of the zenith angle. This calibration error may be significant for HD98800 ($\sim$ 1 to 2 \kms\ ), since it was observed at low elevation angles, where the effect is greatest.

However, offsets in the wavelength calibration do not affect the precision with which the relative velocities of the components can be measured, since the light from each component of the multiple system is displaced by the same amount.
Also, the position angle of the A-B system varied with respect to the slit between nights by 3.6 degrees, and this could, in principle, give rise to variations in the relative separation between the lines of the A and B systems. But this effect is negligible, leading to a relative shift of $\sim 0.2$ \kms.

We also note that the predicted and observed radial velocity separations between Ba and Bb differ systematically by about 6 \kms\ for all three nights, which is significantly larger than expected from our measurement precision (Figure \ref{vradfitfig1}).
It is possible that this is an artifact due to differences in the velocity scales between this work and the earlier radial velocity measurements of T95, on which the predictions are based. However, this would imply an unlikely systematic underestimate of all our radial velocities by about $10\%$ relative to T95. 
As an alternative explanation, we hypothesized that the radial velocity predictions were offset from the true orbit due to inaccuracies in the orbital period of HD 98800 B.

\begin{figure}
\caption{Same as Figure \ref{vradfitfig1}, except by using predicted radial velocities from the ephemeris 4.5 days after the actual observing dates \label{vradfitfig2}}
\plotone{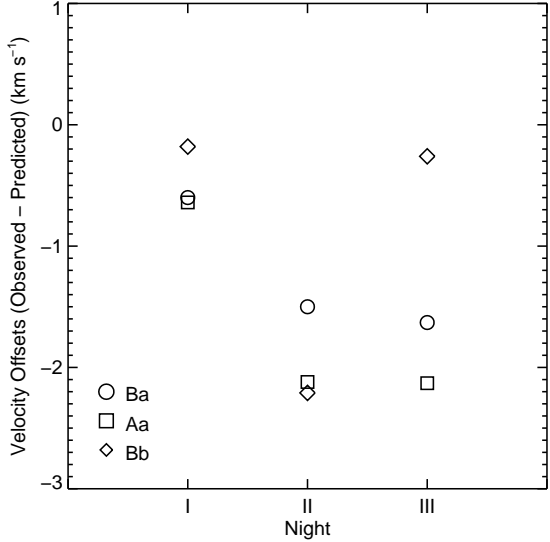}
\end{figure}

To test this, we determined the date when the radial velocity separation between Ba and Bb predicted by T95 was closest to the observed separation. We found that if we used the predictions of T95 for a date 4.5 days later than the date of our observation, then the systematic effect vanished and the rms separation was reduced to 0.9 \kms\ (Figure \ref{vradfitfig2}). This implies an orbital period of $314.89\pm0.03$ days, which is 0.26 days shorter than estimated by T95 and also closer to the value of $314.327\pm0.028$ determined by B05. However, this estimated orbital period is still significantly discrepant from B05. Continual measurements of radial velocities covering a wider range of epochs are needed to adequately resolve this issue. 

\begin{deluxetable*}{c c c c c c c c c c c}
\tabletypesize{\footnotesize}
\tablewidth{0pt}
\tablecaption{Best fit results \label{bestfits}} 
\tablehead{
   \colhead{Night} & \colhead{$T_{\rm Aa}$} & \colhead{$T_{\rm Ba}$} 
 & \colhead{$T_{\rm Bb}$} & \colhead{$R_{\rm Ba}/R_{\rm Aa}$} & \colhead{$R_{\rm Bb}/R_{\rm Aa}$}
 & \colhead{\feh} & \colhead{$c_{\rm Aa}$}
 & \colhead{$c_{\rm Ba}$} & \colhead{$c_{\rm Bb}$} & \colhead{$\chi^2$}\\
  \colhead{ } & \colhead{(K)} & \colhead{(K)} 
 & \colhead{(K)} & \colhead{ } & \colhead{ }
 & \colhead{ } & \colhead{ }
 & \colhead{ } & \colhead{ } & \colhead{ }
}
\startdata
I   & 4535 & 4200 & 3500 & 0.6 & 0.4 & $-0.21$ & 0.686 & 0.229 & 0.085 &  9.2\\
II  & 4535 & 4200 & 3500 & 0.7 & 0.5 & $-0.18$ & 0.607 & 0.276 & 0.117 & 16.0\\
III & 4515 & 4200 & 3500 & 0.7 & 0.5 & $-0.21$ & 0.606 & 0.276 & 0.118 & 15.4\\
\enddata
\end{deluxetable*}

\subsection{Matching observed and modeled spectra}
\label{matching}
VF05 used SME to determine the Na, Si, Ti, Fe and Ni abundances in 1,040 nearby F, G, and K stars. They used spectra spanning wavelengths from 5160 \AA\ to 5190 \AA{} and 6000 \AA to 6180 \AA{}, from the Lick, Keck and AAT planet search programs. As a prerequisite to their analysis, they tuned the atomic line data to match the observed spectrum of the Sun with synthesized solar spectra. We used the same wavelength interval as VF05, except that we avoided the Mg b-band (5160 - 5190 \AA) due to difficulties with continuum normalization in that region. We note that in VF05 this band was used to constrain the stellar gravity, a parameter that we do not fit for in our analysis. Our wavelength range of 6000 -- 6180 \AA{} includes the rest of the range used by VF05.  This enabled us to adopt line parameters from VF05 for computing our model spectra.

VF05 also found that some spectral features were unsuitable for analysis, certain Ca lines in particular being impossible to match over a broad range of temperatures with their models. They thus generated wavelength masks that were used to define the pixels used to compute the $\chi^2$ goodness of fit for their modeled spectra. For fitting the HD 98800 spectrum and determining the $\chi^2$ statistic, we modified these masks to encompass the lines from the three stars with their different radial velocities. This was done by taking the mask for a single star, shifting it in radial velocity for each star of the 98800 spectrum and then combining the three masks together to generate a final mask that included all individual `good' pixels from the three masks. This procedure increases the number of good pixels, but we still avoided the fitting of most `bad' lines, such as the \ion{Fe}{1} line at 6065.482 \AA, the \ion{Fe}{1} lines near 6137 \AA, and the \ion{Ca}{1} 6122.2 line, which has broad damping wings. Full details of the lines that were fitted and the relevant masks in the lab frame (for a single star) can be found in VF05.

We validated our procedure of spectrum fitting using observations of 11 nearby G dwarfs with \teff\ between 5400 and 6000 K that were modeled earlier by VF05 using SME. We observed these stars contemporaneously with our observations of HD 98800 using the same instrument and set up (the Lick Hamilton spectrograph). Using SME, we found that we could match the metallicities found in VF05 within 0.04 dex, which places our results on the same abundance scale as VF05.  Details are given in \citet{Soderblom2009}.

\subsection{Determination of abundances}
\label{abunds}
Having estimated the radial velocities for a given night, we proceeded to fit for the metallicities and temperatures of the three components. To do this, we computed and compared the $\chi^2$ goodness of fit between the modeled spectrum and the observed spectrum for different parameters of the three stars. We varied the temperatures and radii (to calculate the flux ratios: see equation \ref{fluxratioeqn1}) of the three components as well as the overall metallicity of the system over a grid of values, generating one $\chi^2$ value for each possible parameter combination. As an initial grid, we took five temperature steps of 150 K starting at 4200 K for Aa, 3750 K for Ba and 3550 K for Bb. Five steps were also taken for each of the stellar radii centered on the values from B05 with steps equal to the quoted uncertainties ($R_{\rm Aa}$: 1.27, 1.43, 1.59, 1.75, 1.91 $R_\sun$; $R_{\rm Ba}$: 0.67, 0.81, 0.95, 1.09, 1.23 $R_\sun$; $R_{\rm Bb}$: 0.52, 0.63, 0.74, 0.85, 0.96 $R_\sun$). Our selected \feh\ points were $+0.3, 0.0, -0.1, -0.2, -0.3, -0.4$ and $-0.6$ dex. Thus our range of parameter values included the temperatures, radii, and metallicities determined by B05. We then looked for the model with the minimum $\chi^2$ and repeated this process for spectra taken on each of the three nights.

Based on the results of this preliminary analysis with a coarse grid, we adopted a finer grid centered on the best-fit \teff\ and \feh\ values from the coarse-grid analysis. The fine grid had five temperature steps, with $T_{\rm Aa}$ ranging from 4495 to 4575 K (20 K steps), $T_{\rm Ba}$ from 4050 to 4350 K (75 K steps) and $T_{\rm Bb}$ from 3500 to 3800 K (75 K steps). We also fixed the radius of Aa to unity and normalized the other radii with respect to $R_{\rm Aa}$, taking both $R_{\rm Ba}/R_{\rm Aa}$  and $R_{\rm Bb}/R_{\rm Aa}$ from 0.0 to 1.0, thereby obtaining flux ratios for Aa:Ba:Bb from 1:0:0 to approximately 0.33:0.33:0.33. From the coarse grids, it was clear that an \feh\ value near $-0.2$ dex produced the minimum $\chi^2$ values, and so we took \feh\ from $-0.27$ to $-0.12$ dex, spaced at 0.03 dex. The parameter combinations giving the minimum $\chi^2$ for spectra taken on each of the three nights with this fine grid are shown in Table \ref{bestfits}.

The $T_{\rm Bb}$ values favoring the best fit are at the lowest temperature in our grid of model atmospheres (3500 K). Our estimate for $T_{\rm Bb}$ is also lower than the value (4000 K) in B05. We note that the few lines we had to work with for Bb only provided weak constraints on its temperature. Therefore we adopt 3650 K, which is the next available grid point in temperature for Bb from the optimal-$\chi^2$ solution at 3500 K, as an estimate of its temperature. Further, we find that the best fit values of $T_{\rm Aa}$ for a given \feh\ are correlated with \feh\ itself. This can be understood by the fact that the Aa component has a dominant contribution to the spectrum and that increasing metallicity increases the depths of spectral lines, which is offset in these fits by a rising best fit temperature for the dominant component. The error in the determination of metallicity for HD 98800 is dominated by systematic effects and is of the order of $0.1$ dex, the origin of which is discussed in \citet{Soderblom2009}. This is not a measure of the internal error in this analysis, which is $\sim 0.03$ dex (from Table \ref{bestfits}) and negligible compared to the systematics.

\begin{deluxetable}{r c c c c c c}
\tablewidth{0pt}
\tablecaption{Derived relative radii versus \feh \label{photofit}}
\tablehead{
\colhead{ } & \colhead{$T_{\rm Aa}$} & \colhead{$T_{\rm Ba}$} & \colhead{$T_{\rm Bb}$} & 
\colhead{$ $} & \colhead{$ $} & \colhead{$ $}\\
\cline{2-4}
\colhead{\feh} & \colhead{(K)} & \colhead{(K)} & \colhead{(K)} & 
\colhead{$R_{\rm Ba}/R_{\rm Aa}$} & \colhead{$R_{\rm Bb}/R_{\rm Aa}$} & \colhead{$\chi^{2}$}
}
\startdata
$-0.6$ &   4350 &    4050 &    3500 &    0.71 &   0.32 &  16.5\\
$-0.4$ &   4350 &    4200 &    3500 &    0.71 &   0.32 &  12.3\\
$-0.3$ &   4500 &    3900 &    3500 &    0.80 &   0.36 &  11.4\\
$-0.2$ &   4500 &    4350 &    3500 &    0.80 &   0.36 &  11.3\\
$-0.1$ &   4650 &    3750 &    3500 &    0.80 &   0.36 &  11.8\\
 $0.0$ &   4650 &    3750 &    3500 &    0.80 &   0.36 &  12.5\\
\enddata
\end{deluxetable} 

\subsection{Flux Ratios and photometry matching}

\label{photomatching}
We find that the line ratios for the three stars change from night to night, with the lines for Ba becoming relatively stronger from the first to the second night and the lines from Aa becoming relatively weaker. We infer (Section \ref{vradtext}) that the stars may not have been well placed in the center of the spectrograph slit during the observations and hence there is some uncertainty in the determination of the true flux ratios (at $\sim 6090$ \AA) of the stars and hence in their radii. However, this does not affect the ratio Bb/Ba. We find the mean and standard deviations of this ratio for the lowest 40 $\chi^2$ values for the three observing nights to be $0.376\pm0.007$ $(\Delta\chi^2=0.12)$, $0.430\pm 0.007$ $ (\Delta\chi^2=0.23)$ and $0.435\pm0.010$ $(\Delta\chi^2=0.20)$. This corresponds to a SNR-weighted Bb/Ba average flux ratio of $0.416\pm0.005$. The result does not depend strongly on the number of low-$\chi^2$ solutions averaged.

\textit{HST} WFPC2 photometry for the system carried out by \citet{Soderblom1998} yielded flux ratios for the components with Aa:Ba:Bb::0.62:0.316:0.064 in the $V$ band, with Bb/Ba of $0.207$. We carried out a set of runs by fixing the flux ratios of the three stars at these values and fitting the spectrum on the first night for temperatures and metallicities. The results are shown in Table \ref{photofit}. Using flux ratios based on the photometry alone, we find that the best fits favor sub-solar metallicity for the system. A value of $\feh\ \sim -0.2$ to $-0.3$ is indicated from these fits.

\section{Conclusions}
We have fitted high resolution spectra of the HD 98800 system to determine an overall metallicity of $-0.20\pm0.10$. This confirms indications by \citet{Boden2005} that HD 98800 is a system with lower metal abundance than the Sun. Coupled with age estimates of HD 98800 placing it between 8 and 20 Myrs, this would imply that there is ongoing star formation in the solar neighborhood forming relatively metal-poor systems even in the present epoch. This result is consistent with \citet{Santos2008}, who find $-0.13 \le $ [{\rm {Fe/H}] $\le -0.05$ for six nearby star-forming regions and with \citet{James2006} who find  $-0.17 \le$ [{\rm {Fe/H}] $\le -0.02$ for three southern star-forming regions.

We have also estimated the effective temperatures of the visible components to be $T_{\rm Aa} = 4530 \pm 80$ K (stat), $T_{\rm Ba} = 4200 \pm 90 $ K (stat), $T_{\rm Bb} \sim 3650$ K and the flux ratio for Bb/Ba to be $0.416\pm0.005$ at 6090 \AA. Finally, we have determined a new orbital period of $314.89\pm0.03$ days for the B-system.

We note that it is possible for the apparent metallicity derived from a stellar spectrum to be lowered in the presence of very high activity levels, due to filling in of line cores.  The HD 98800 system is very young, but the activity seen in its components is modest.  For example, none of the components exhibits H$\alpha$ emission strong enough to rise above the continuum.  The activity seen in HD 98800 is also less than in Pleiades members of similar color, and, as we show in \citet{Soderblom2009}, we can measure metallicity in those Pleiads without any apparent effect from activity.  In particular, the abundances show no trend with color in going from F dwarfs into G stars in the Pleiades \citep{Soderblom2009}, despite a systematic rise in activity levels.  We conclude that our determination of metallicity in HD 98800 is unlikely to be affected significantly by stellar activity.

One potential source of systematic error which could cause our metallicity estimate to be biased towards low metallicity is our assumption that the faint fourth companion Ab contributes no light to our spectra.  If, instead, Ab had significant flux in our wavelength range (and particularly if it were a rapid rotator), it would raise the continuum level and hence reduce the measured equivalent widths of all lines - which our modeling would interpret as indicative of a lower metallicity.  Because Ab has been detected now in high-resolution imaging (B05), we can estimate the effect of Ab on our spectra. Boden (private communications, 2008) estimates that at the \textit{K} band, the Ab/Aa contrast ratio is 15\%. From the measured  near-IR photometry for HD98800 and a nominal distance of 40 pc (B05), we therefore estimate absolute magnitudes of \textit{K}(Aa) $= 3.4$ and \textit{K}(Ab) $= 5.5$. Using the \citet{Siess2000} 10 Myr isochrones, these absolute K  magnitudes correspond to spectral types about K6 and M6. The Siess isochrones also provide estimated broadband $R-K$ colors, with $V-R \sim 2.3$ for K6 and $V-R \ge 5.7$ for M6, yielding R-band magnitudes of $R$(Aa) $ = 5.7$ and $R$(Ab) $= 11.3$. This indicates that the flux ratio of Ab/Aa in the wavelength range from which we derive metallicities is less than 1\%; the influence of Ab on the combined spectrum including Ba and Bb is even less. We therefore conclude that excluding Ab from our analysis has a negligible (less than 1\%) influence on our derived metallicity.

The stellar parameters we determine could be checked for consistency with pre-main sequence stellar models by computing evolutionary tracks with the above parameters and checking whether the stars truly lie on a single isochrone. Fully resolved separate spectra of the components of HD 98800 would provide more accurate parameters, which should be feasible with adaptive optics systems given the separations of the stars. Astrometry using adaptive optics would also improve estimates of dynamical masses for the components, enabling tighter constraints to be placed on the age of the system. HD 98800 will continue to be an important reference system for understanding pre-main sequence evolution.

\acknowledgments
We thank G. Torres of the Center for Astrophysics for providing information on the orbits of the components of HD 98800
and A. F. Boden of the California Institute of Technology for helpful discussions.
 This work was done under the auspices of the Summer Student Program at the Space Telescope Science Institute.

\bibliography{citations}

\begin{thebibliography}{25}
\expandafter\ifx\csname natexlab\endcsname\relax\def\natexlab#1{#1}\fi

\bibitem[{{Alencar} \& {Batalha}(2002)}]{Alencar2002}
{Alencar}, S.~H.~P. \& {Batalha}, C. 2002, \apj, 571, 378

\bibitem[{{Baraffe} {et~al.}(1998){Baraffe}, {Chabrier}, {Allard}, \&
  {Hauschildt}}]{Baraffe1998}
{Baraffe}, I., {Chabrier}, G., {Allard}, F., \& {Hauschildt}, P.~H. 1998, \aap,
  337, 403

\bibitem[{{Boden} {et~al.}(2005){Boden}, {Sargent}, {Akeson}, {Carpenter},
  {Torres}, {Latham}, {Soderblom}, {Nelan}, {Franz}, \&
  {Wasserman}}]{Boden2005}
{Boden}, A.~F., {Sargent}, A.~I., {Akeson}, R.~L., {Carpenter}, J.~M.,
  {Torres}, G., {Latham}, D.~W., {Soderblom}, D.~R., {Nelan}, E., {Franz},
  O.~G., \& {Wasserman}, L.~H. 2005, \apj, 635, 442, (B05)

\bibitem[{ESA(1997)}]{ESA1997b}
ESA. 1997, The Hipparcos and Tycho Catalogues (Paris: ESA), SP-1200

\bibitem[{Innes(1909)}]{Innes1909}
Innes, R. 1909, Transvaal Obs. Circ., 1, 1

\bibitem[{{James} {et~al.}(2006){James}, {Melo}, {Santos}, \&
  {Bouvier}}]{James2006}
{James}, D.~J., {Melo}, C., {Santos}, N.~C., \& {Bouvier}, J. 2006, \aap, 446,
  971

\bibitem[{{Kastner} {et~al.}(1997){Kastner}, {Zuckerman}, {Weintraub}, \&
  {Forveille}}]{Kastner1997}
{Kastner}, J.~H., {Zuckerman}, B., {Weintraub}, D.~A., \& {Forveille}, T. 1997,
  Science, 277, 67

\bibitem[{{Kurucz}(1992)}]{Kurucz1992}
{Kurucz}, R.~L. 1992, in IAU Symposium, Vol. 149, The Stellar Populations of
  Galaxies, ed. B.~{Barbuy} \& A.~{Renzini}, 225

\bibitem[{{Kurucz}(1993)}]{Kurucz1993a}
{Kurucz}, R.~L. 1993, ATLAS9 Stellar Atmosphere Programs and 2 km/s
  grid.~Kurucz CD-ROM No.~13.~ Cambridge, Mass.: Smithsonian Astrophysical
  Observatory, 1993., 13

\bibitem[{{Muzerolle} {et~al.}(2000){Muzerolle}, {Calvet}, {Brice{\~n}o},
  {Hartmann}, \& {Hillenbrand}}]{Muzerolle2000}
{Muzerolle}, J., {Calvet}, N., {Brice{\~n}o}, C., {Hartmann}, L., \&
  {Hillenbrand}, L. 2000, \apjl, 535, L47

\bibitem[{{Padgett}(1996)}]{Padgett1996}
{Padgett}, D.~L. 1996, \apj, 471, 847

\bibitem[{{Prato} {et~al.}(2001){Prato}, {Ghez}, {Pi{\~n}a}, {Telesco},
  {Fisher}, {Wizinowich}, {Lai}, {Acton}, \& {Stomski}}]{Prato2001}
{Prato}, L., {Ghez}, A.~M., {Pi{\~n}a}, R.~K., {Telesco}, C.~M., {Fisher},
  R.~S., {Wizinowich}, P., {Lai}, O., {Acton}, D.~S., \& {Stomski}, P. 2001,
  \apj, 549, 590

\bibitem[{{Santos} {et~al.}(2008){Santos}, {Melo}, {James}, {Gameiro},
  {Bouvier}, \& {Gomes}}]{Santos2008}
{Santos}, N.~C., {Melo}, C., {James}, D.~J., {Gameiro}, J.~F., {Bouvier}, J.,
  \& {Gomes}, J.~I. 2008, \aap, 480, 889

\bibitem[{{Siess} {et~al.}(2000){Siess}, {Dufour}, \& {Forestini}}]{Siess2000}
{Siess}, L., {Dufour}, E., \& {Forestini}, M. 2000, \aap, 358, 593

\bibitem[{{Soderblom} {et~al.}(1996){Soderblom}, {Henry}, {Shetrone}, {Jones},
  \& {Saar}}]{Soderblom1996}
{Soderblom}, D.~R., {Henry}, T.~J., {Shetrone}, M.~D., {Jones}, B.~F., \&
  {Saar}, S.~H. 1996, \apj, 460, 984

\bibitem[{{Soderblom} {et~al.}(1998){Soderblom}, {King}, {Siess}, {Noll},
  {Gilmore}, {Henry}, {Nelan}, {Burrows}, {Brown}, {Perryman}, {Benedict},
  {McArthur}, {Franz}, {Wasserman}, {Jones}, {Latham}, {Torres}, \&
  {Stefanik}}]{Soderblom1998}
{Soderblom}, D.~R., {King}, J.~R., {Siess}, L., {Noll}, K.~S., {Gilmore},
  D.~M., {Henry}, T.~J., {Nelan}, E., {Burrows}, C.~J., {Brown}, R.~A.,
  {Perryman}, M.~A.~C., {Benedict}, G.~F., {McArthur}, B.~J., {Franz}, O.~G.,
  {Wasserman}, L.~H., {Jones}, B.~F., {Latham}, D.~W., {Torres}, G., \&
  {Stefanik}, R.~P. 1998, \apj, 498, 385, (S98)

\bibitem[{{Soderblom} {et~al.}(2009){Soderblom}, {Laskar}, {Valenti}, \&
  {Stauffer}}]{Soderblom2009}
{Soderblom}, D.~R., {Laskar}, T., {Valenti}, J.~A., \& {Stauffer}, J. 2009,
  \aj, submitted

\bibitem[{{Tokovinin}(1999)}]{Tokovinin1999}
{Tokovinin}, A.~A. 1999, Astronomy Letters, 25, 669

\bibitem[{{Torres} {et~al.}(1995){Torres}, {Stefanik}, {Latham}, \&
  {Mazeh}}]{Torres1995}
{Torres}, G., {Stefanik}, R.~P., {Latham}, D.~W., \& {Mazeh}, T. 1995, \apj,
  452, 870, (T95)

\bibitem[{{Valenti} \& {Fischer}(2005)}]{Valenti2005}
{Valenti}, J.~A. \& {Fischer}, D.~A. 2005, \apjs, 159, 141, (VF05)

\bibitem[{{Valenti} \& {Piskunov}(1996)}]{Valenti1996}
{Valenti}, J.~A. \& {Piskunov}, N. 1996, \aaps, 118, 595

\bibitem[{van Leeuwen(2007{\natexlab{a}})}]{Leeuwen2007}
van Leeuwen, F. 2007{\natexlab{a}}, Astrophysics and Space Science Library,
  Vol. Volume 350, Hipparcos, the New Reduction of the Raw Data
  (http://www.springer.com/series/5664: Springer Netherlands), 449

\bibitem[{van Leeuwen(2007{\natexlab{b}})}]{Leeuwen2007a}
---. 2007{\natexlab{b}}, \aap, 474, 653

\bibitem[{{Vogt}(1987)}]{Vogt1987}
{Vogt}, S.~S. 1987, \pasp, 99, 1214

\bibitem[{{Yang} {et~al.}(2005){Yang}, {Johns-Krull}, \& {Valenti}}]{Yang2005}
{Yang}, H., {Johns-Krull}, C.~M., \& {Valenti}, J.~A. 2005, \apj, 635, 466

\end{thebibliography}
\bibliographystyle{apj}

\end{document}